\documentclass[aps,
prb,
preprint,
floatfix,
superscriptaddress,
showpacs
]{revtex4-1}

\usepackage{graphicx}
\usepackage{dcolumn}
\usepackage{amsmath,bm}
\usepackage{color}%
\usepackage{multirow}
\usepackage{array} 
\usepackage{esvect}
\usepackage{xr}
\usepackage{braket}
\usepackage{soul}

\usepackage[utf8]{inputenc}
\usepackage{kotex}
\usepackage{kotex-logo}

\begin{document}


\title{Establishing Epitaxial Connectedness in Multi-Stacking: The Survival of Thru-Holes in Thru-Hole Epitaxy} 

\author{Youngjun Lee}
\affiliation{Department of Physics, and Research Institute for Basic Sciences, Kyung Hee University, Seoul 02447, Korea}

\author{Seungjun Lee}
\altaffiliation[Current address: ]{Department of Electrical and Computer Engineering, University of Minnesota, Minneapolis, Minnesota 55455, USA}
\affiliation{Department of Physics, and Research Institute for Basic Sciences, Kyung Hee University, Seoul 02447, Korea}

\author{Jaewu Choi}
\affiliation{Department of Information Display, Kyung Hee University, Seoul 02447, Korea}
 
\author{Chinkyo Kim}
\email[Corresponding author. E-mail: ]{ckim@khu.ac.kr}
\affiliation{Department of Physics, and Research Institute for Basic Sciences, Kyung Hee University, Seoul 02447, Korea}
\affiliation{Department of Information Display, Kyung Hee University, Seoul 02447, Korea}
 
\author{Young-Kyun Kwon}%
\email[Corresponding author. E-mail: ]{ykkwon@khu.ac.kr}
\affiliation{Department of Physics, and Research Institute for Basic Sciences, Kyung Hee University, Seoul 02447, Korea}
\affiliation{Department of Information Display, Kyung Hee University, Seoul 02447, Korea}

\date{\today}

\begin{abstract}
Thru-hole epitaxy has recently been reported to be able to grow readily detachable domains crystallographically aligned with the underlying substrate over 2D mask material transferred onto a substrate. [Jang \textit{et al.}, \textit{Adv. Mater. Interfaces}, \textbf{2023} \textit{10}, 4 2201406] While the experimental demonstration of thru-hole epitaxy of GaN over multiple stacks of $h$-BN was evident, the detailed mechanism of how small holes in each stack of $h$-BN survived as thru-holes during multiple stacking of $h$-BN was not intuitively clear. Here, we use Monte Carlo simulations to investigate the conditions under which holes in each stack of 2D mask layers can survive as thru-holes during multiple stacking. If holes are highly anisotropic in shape by connecting smaller holes in a particular direction, thru-holes can be maintained with a high survival rate per stack, establishing more epitaxial connectedness. Our work verifies and supports that thru-hole epitaxy is attributed to the epitaxial connectedness established by thru-holes surviving even through multiple stacks.
\end{abstract}


\maketitle

\section{Introduction}

Remote epitaxy has attracted much attention because of its fascinating epitaxial growth of readily detachable crystalline domains, which are crystallographically aligned with the underlying substrate over 2D material without any form of direct bonding between the material grown and the substrate.\cite{Kim_Nature_544_340,Kong_NM_17_999,Jeong_SA_6_eaaz5180}  It was argued that remote epitaxial growth was possible because crystallographic information of the substrate is well transferred through ultrathin 2D layers especially if the underlying substrate has a strong ionic or polar character. Crystalline films without direct chemical bonding with the substrate are consequently shown to be readily detached.  Although remote epitaxy opened a new possibility for the growth of 3D materials with barely any constraints to the lattice match with the substrate, it is still not straightforward to carry out remote epitaxy because it requires stringent conditions: (i) defect-free 2D layers, (ii) precise controllability for the number of 2D layers, and (iii) polar characters of materials to be grown, to name a few.\cite{Kim_Nature_544_340,Kong_NM_17_999,Jeong_SA_6_eaaz5180} While current 2D material growth and transfer techniques have greatly improved, it is still difficult or impossible to transfer completely defect-free graphene onto a substrate.

It is evident that not only do 2D materials prepared by current techniques contain structural defects such as holes and small cracks but they can also be degraded. Thus, during its transfer process, such structural defects would be extended in a specific direction.\cite{RAKIB201767,Budarapu2015} Because the substrate is well exposed as a consequence, such a degraded 2D material will act as a mask for epitaxial lateral growth rather than as a transparent overlayer of substrate potential for remote epitaxy.\cite{Heilmann2015, He2017, Fernandez2017}. Furthermore, it can be inferred that even the film grown through holes in 2D material can also be readily detached since the 2D material mostly acts as a mask over a crystalline substrate. We indeed showed that this can be possible without those stringent conditions required for remote epitaxy by demonstrating not only the growth of crystallographically aligned GaN domains over a 2D material, such as $h$-BN, but also the facile detachment of those grown and merged GaN domains.\cite{thruhole} This epitaxial approach was named thru-hole epitaxy. Note that the term ``thru-hole'' is used to refer to a hole connected all the way from the top-most 2D material to the substrate resulting in the establishment of the epitaxial connectedness between the material grown and the substrate. The same thru-hole epitaxy was also successfully applied to grow GaN and ZnO over graphene and MoS$_2$, respectively.\cite{Lee2022} There was another report showing epitaxial growth of GaSb films on graphene-terminated surface by a similar approach called pinhole-seeded lateral epitaxy.\cite{Manzo2022} 

\begin{figure}[t]
\centering
\includegraphics[width=1.0\columnwidth]{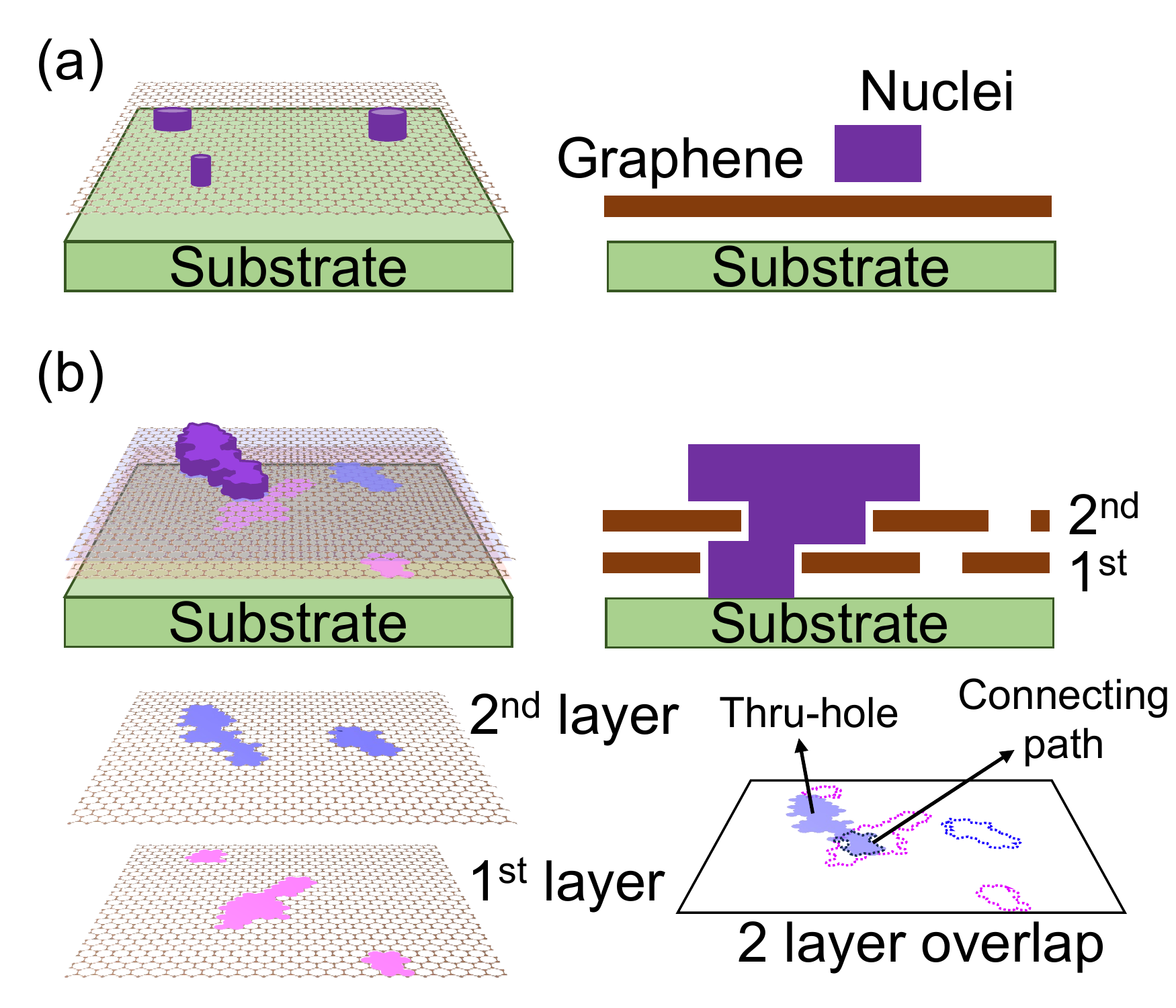}
\caption{Schematic illustration of (a) remote epitaxy and (b) thru-hole epitaxy occurring on a 2D material transferred onto a substrate. In the former process, nucleation occurs on a defect-free 2D monolayer, while in the latter process, nucleation occurs directly on the substrate after diffusion via thru-holes in a bilayer with some defects. In (b), the areas colored in blue and red represent defects or holes in the first and second layers, respectively. This bilayer has a maze of thru-holes from the top surface all the way down to the substrate .}
\label{thruhole}
\end{figure}

The growth mechanisms of remote and thru-hole epitaxy are completely different. This distinction stems from whether or not there are thru-holes, through which atomic species adsorbed on the 2D material can diffuse into the substrate, as illustrated in Fig.~\ref{thruhole}. In remote epitaxy, the substrate potential should be sneaked out through the 2D material because there are no thru-holes (see Fig.~\ref{thruhole}(a)). In thru-hole epitaxy, on the other hand, the adsorbed atoms on the 2D material can permeate into the substrate via thru-holes and form the nuclei directly on the substrate, as displayed in Fig.~\ref{thruhole}(b). Unlike remote epitaxy, which stringently requires a defect-free graphene monolayer, we observed that thru-hole epitaxy can be successfully carried out with tens of nanometers thick (multiply-stacked) $h$-BN layers or even thicker SiO$_2$ as long as thru-holes are maintained.\cite{thruhole} Here, a `stack' represents a transfer unit consisting of multiple layers of $h$-BNs grown together at the same time. Crystallographically-aligned GaN domains were successfully grown over multiply-stacked $h$-BN transferred onto the sapphire substrate and easily detached with a thermal release tape. Despite this clear and evident experimental demonstration of thru-hole epitaxy, there remains still one important issue, which is not intuitively convincing. How can small holes in each layer or stack of $h$-BN survive as thru-holes without being sealed during multiple stacking? This question naturally arises because our intuition, at first sight, would say that the probability of the survival of holes as thru-holes in multiple stacking would quickly drop close to zero even by a few times of stacking of 2D materials containing a moderate concentration of structural defects.

This is a critical question to answer in order to validate thru-hole epitaxy in multiply-stacked 2D materials because, in thru-hole epitaxy, the epitaxial connectedness between the film to be grown and the underlying substrate can only be established by thru-holes. In other words, if such holes are completely sealed in multiple stacking, thru-hole epitaxy will not occur. However, the successful observation of the thru-hole epitaxy of GaN on eight-time stacked $h$-BN\cite{thruhole} remained puzzling enough that we had to look for convincing evidence that holes survive as thru-holes even with multiple stacks. If the holes in each stack are isolated from each other, we can expect the number of thru-holes to decrease exponentially with the number of stacks. However, if the holes in each stack are not well isolated, the decrease rate of thru-holes with the number of stacks can be reduced. The $h$-BN samples used in our previous study were fabricated without a growth optimization process, so it is likely that structural defects such as cracks were present.\cite{thruhole} These cracks in particular can act as laterally connected holes rather than isolated ones. Furthermore, thru-holes do not need to be vertically straight but can be meandering and crooked like a maze of connecting paths, as shown in Fig.~\ref{thruhole}(b), as long as they simply run all the way from the top to the bottom stack, and thus to the substrate, to establish epitaxial connectedness.

In this study, we present the results of Monte Carlo simulations that show how holes present in each stack of a 2D material survive as thru-holes during multiple stacking and how their survival probability depends on the hole configuration in the 2D material. In our Monte Carlo simulations, we considered different hole configurations, such as isolated holes, short connected holes, linearly connected holes, and so on. We also evaluated the survival rate of thru-holes while varying the areal fraction, or concentration, of holes in each 2D stack. The results show that in certain configurations where holes are more like long and narrow in shape, more holes survive as thru-holes during multiple stacks, even at low hole concentrations, confirming our previous experimental observations.\cite{thruhole}

\section{Method}
\label{method}

\begin{figure}[t]
\centering
\includegraphics[width=1.0\columnwidth]{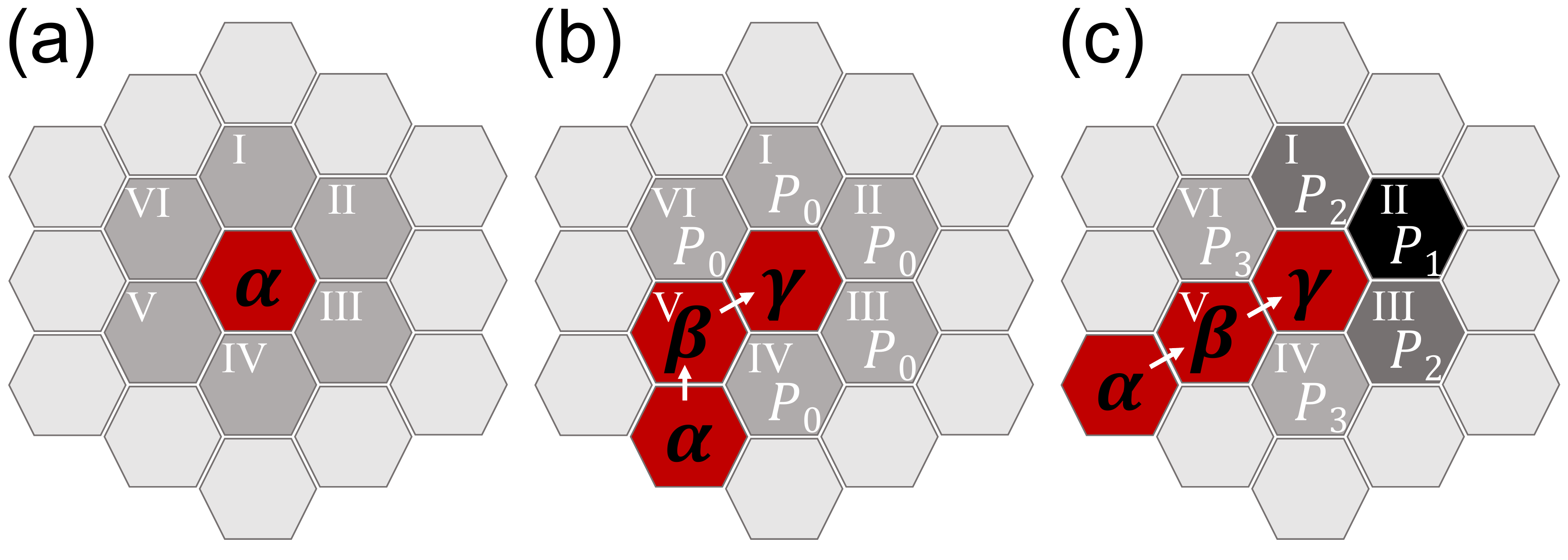}
\caption{\label{size-of-hole} Construction schematics of different hole configurations. (a) Size variations: (HC1) a hole with one randomly selected hexagonal unit cell $\alpha$ open, (HC2) a hole with $\alpha$ as in (HC1) and three additional cells selected from the six gray-shaded nearest neighboring cells, I$\sim$VI, of $\alpha$, and (HC3) a hole with $\alpha$ and all six neighboring cells (I$\sim$VI) selected to be open. Note that a hole in (HC2) can have a different shape depending on the selection of the nearest neighboring cells, e.g., a hole composed of $\alpha$ and three additional cells (I, II, and III), (I, III, and IV), or (I, III, V). Note also that even in the case of (HC1), a hole consisting of two or more connected hexagonal cells can exist, because, although it is very small, there is still a chance of selecting an adjacent cell on the next trial.
(b) Shape variations: Suppose that three cells $\alpha$, $\beta$, and $\gamma$ are selected as open in sequence. The next open cell is selected with probability $P_0$ from the five gray-shaded cells surrounding the $\gamma$ cell (I$\sim$VI, excluding V, which is the already selected $\beta$ cell). The same process was used when the $\gamma$ cell was selected from the $\beta$ cell. Note that there is still a non-zero probability of $1-5P_0$ that an independent cell other than the surrounding cells will be selected as open, and so this process is repeated until either an independent cell or one of the already selected open cells is selected. In our simulations, we considered two cases of (HC4) and (HC5) with $P_0=0.14$ and $P_0=0.18$, respectively.
(c) Directionality variations to mimic a realistic crack propagation: Once three linearly connected hexagonal cells $\alpha$, $\beta$, and $\gamma$ have been sequentially and consecutively chosen open, a next open cell is chosen among five surrounding cells with different probabilities $P_1>P_2>P_3$, weighted by their straightness. In our simulations, we considered two cases of (HC6) and (HC7) with $(P_1,P_2,P_3)=(0.45,0.18,0.045)$ and $(P_1,P_2,P_3)=(0.63,0.09,0.045)$, respectively. The scale of the gray shading in the cells is an indication of directionality. It is clear that (HC7) will produce more straight-looking holes than (HC6). Note that there is a probability of 0.1 that a cell other than the surrounding cells (I$\sim$VI) is selected as open.}
\end{figure}

To more realistically estimate the survival ratio of holes that eventually become part of a thru-hole during $h$-BN stacking, we performed Monte Carlo simulations while varying the distribution of holes and the configuration of individual holes in each stack of 2D material. A monolayer or single stack of 2D material was modeled as consisting of a $1000\times1000$ grid of hexagonal unit cells, as partially shown in Fig.~\ref{size-of-hole}. Each hexagonal unit cell can be either open or closed. Each open cell represents a hole or a part of a larger hole. These hexagonal unit cells do not correspond to the crystallographic unit cells of a hexagonal 2D material, but rather to a minimum size hole through which atoms can diffuse, so they can be part of a thru-hole. Therefore, a larger hole formed by connected open hexagonal cells can act as part of a thru-hole, which is better for diffusing atomic species into the substrate. We considered several different distributions of holes in a single stack or monolayer of 2D material by categorizing holes into seven configurations as described in Fig.~\ref{size-of-hole}. We then repeated the simulation with these seven configurations to see how the size and shape of the holes affect the areal fraction of thru-holes $A_\mathrm{th}$ as the number of stacks $n$ increases. The areal ratio of thru-holes is evaluated as follows: (i) counting the number of open cells in the top stack that connect through all stacks to the bottom substrate, and (ii) dividing this number by $1000\times1000$, the total number of hexagonal cells in the model stack. Note that, as mentioned above, holes do not have to be connected vertically in a straight line to survive as part of a thru-hole. Instead, they can be meandering and crooked, like a connected path in a maze. To evaluate how the survival probability of a thru-hole depends on the configuration of holes and the areal fraction of holes $a$ created in each stack, we ran Monte Carlo simulations while independently varying these two conditions. For each set of these two conditions, we estimated the corresponding thru-hole areal fraction value ($A_\mathrm{th}$) by taking the ensemble average over 1600 samples. We also used the breadth-first Search method to accurately count thru-holes without missing any.~\cite{thomas2016introduction}

Figure~\ref{size-of-hole} shows the construction schematics of seven different hole configurations, (HC1) through (HC7), in each stack. The first three configurations, (HC1), (HC2), and (HC3) represent the size variation of a single hole as shown in Fig.~\ref{size-of-hole}(a). Suppose a hexagonal cell $\alpha$ was randomly chosen to be open at the beginning. For (HC1), the next open cell was then chosen completely independently. For (HC2) and (HC3), three and six cells, respectively, were selected as open among the six nearest neighboring hexagonal cells I$\sim$VI surrounding the $\alpha$ cell. The sizes of most holes in these configurations correspond to those of one, four, and seven hexagonal cells, respectively. Note, of course, that there is still a chance that larger holes can be formed by merging two or more independently generated holes. Note also that four-cell-sized holes generated in (HC2) may have different shapes depending on the choice of the three nearest neighboring cells, while those in (HC3) have the same shape.

\section{Results and discussion}

\begin{figure}
\centering
\includegraphics[width=1.0\columnwidth]{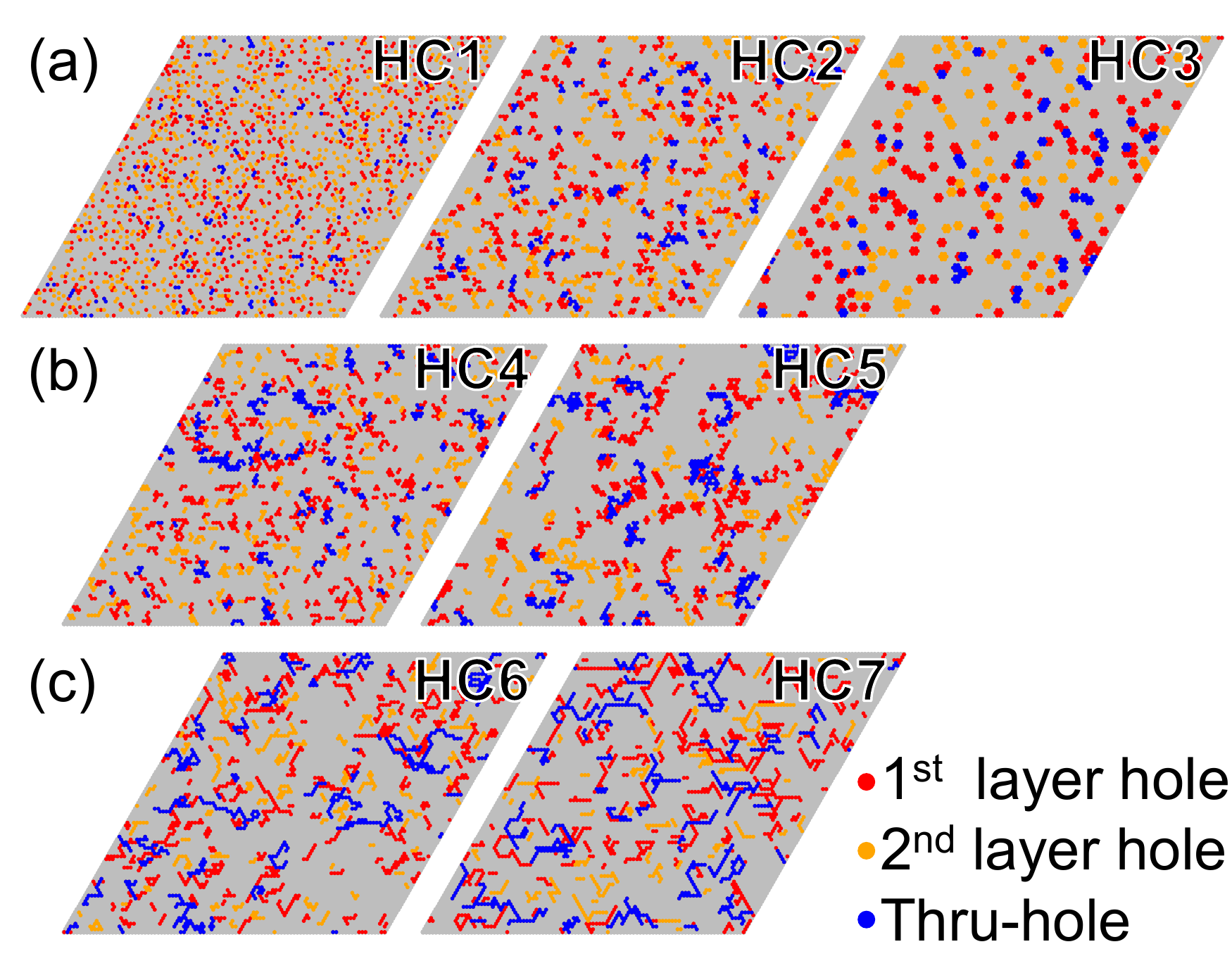}
\caption{\label{simulationI} Thru-hole formation through two layers (or stacks) with (a) (HC1), (HC2), and (HC3), (b) (HC4) and (HC5), and (c) (HC6) and (HC7). In each layer, holes were created to have the same areal fraction of $a=0.1$ or 10\% for demonstration and comparison. The red and orange colors represent holes present in the first and second layers, respectively, while the blue color indicates thru-holes that survived after stacking two layers. It is noted that there are more overlapped holes or thru-holes in (c) than in (a) and (b), indicating that more thru-holes survive through stacks with a directional hole configuration, such as (HC6) or (HC7).}
\end{figure}

To explore how the holes in each layer (stack) survive as part of thru-holes after stacking multiple layers, we first prepared a single layer or stack, with holes created according to one of seven hole configurations (HC1) to (HC7), described above, with a specific hole areal fraction $a$ ($\in[0.01, 0.15]$), then, placed another layer prepared with the same hole configuration and $a$ on top of the first layer. This process was repeated until the number of stacks reached a target number of stacks, $N~(\in[1,10])$. For example, Fig.~\ref{simulationI} demonstrates thru-hole formation in double-layered ($N=2$) 2D material with seven different hole configurations (HC1) to (HC7) and with $a=10$\%. In the figure, holes in the first or bottom layer and the second or top layer are colored red and orange, respectively. A thru-hole that guarantees epitaxial connectedness can be established as long as a hole in the second layer overlaps, even partially, with another hole in the first layer. These thru-holes are colored blue in Fig.~\ref{simulationI}.

In Fig.~\ref{simulationI}(a), the average sizes of thru-holes were estimated to be 1.89, 7.40, and 11.18 hexagonal cells in the respective hole configurations (HC1), (HC2), and (HC3) described in Sect.~\ref{method} and Fig.~\ref{size-of-hole}(a), indicating that the average size of thru-holes increases with the average size of individual holes in each stack, which is reasonable to understand. In the case of (HC1) with two layers, the areal fraction of surviving thru-holes was estimated to be $A_\mathrm{th}^\mathrm{(HC1)}=0.017$, which is quite close to the intuitive estimate of $a^2=0.01$. Note that the intuitive estimate is based on an ideal situation where each hole in each layer is a single hexagonal cell completely isolated from the others. For (HC2) and (HC3), on the other hand, it was estimated to be $A_\mathrm{th}^\mathrm{(HC2)}=0.034$ and $A_\mathrm{th}^\mathrm{(HC3)}=0.0314$, respectively. However, there are two critical surprises in the trend of the areal fraction of surviving thru-holes in (HC1)$\sim$(HC3). The first surprise is that the areal fractions of thru-holes for (HC1)$\sim$(HC3) are different, even though the areal fractions of holes in each stack of (HC1)$\sim$(HC3) were set equal to $a=0.1$. This is because the survival probability of holes as thru-holes during stacking increases with hole size. The second surprise is that $A_\mathrm{th}^\mathrm{(HC2)}=0.034$ is slightly larger than $A_\mathrm{th}^\mathrm{(HC3)}=0.0314$, although the average hole size in (HC3) is larger than that in (HC2). To understand this surprise, we first noticed a difference in the shape anisotropy or aspect ratio of the holes between the two configurations. Holes composed of four hexagonal cells in (HC2) mostly have shape anisotropy, as shown in Fig.~\ref{size-of-hole}(a), while those composed of seven hexagonal cells in (HC3) form an isotropic hexagon. It is the shape anisotropy that caused this surprise. Suppose we can roughly simplify the holes as ellipses. Some holes in (HC2) are modeled by an ellipse with a semi-major axis equal to the radius of a circle representing holes in (HC3). Therefore, there is a possibility that holes in (HC2) overlap with holes in the other layer within the same spatial range as holes in (HC3) do,  even though the former are on average 4/7 times smaller in size than the latter. Moreover, since the areal fractions of holes created in both (HC2) and (HC3) are the same at $a=10$\%, there are about 7/4 times more holes in (HC2) than in (HC3), increasing the chance that a hole in one layer overlaps with holes in an adjacent layer in (HC2). As a result, the areal fraction of thru-holes became slightly larger in (HC2) than in (HC3). This shape anisotropy effect on thru-hole survival during multiple stacking becomes even more pronounced in the other hole configurations, (HC4) through (HC7), as discussed below.

In the case of Fig.~\ref{simulationI}(b), corresponding to the hole configurations (HC4) and (HC5), in which holes were created by selecting hexagonal cells to be open with two different weighted probabilities of $P_0$ as described in Sect.~\ref{method} and Fig.~\ref{size-of-hole}(b). Our simulations revealed that the average sizes of surviving thru-holes in (HC4) and (HC5) were estimated to be 9.05 and 21.9 hexagonal cells, respectively, and $A_\mathrm{th}^\mathrm{(HC4)}=0.039$ and $A_\mathrm{th}^\mathrm{(HC5)}=0.052$. As shown in Fig.~\ref{simulationI}(b), not only did the average areas of the holes increase, but the holes also became more anisotropic in shape than those in (HC1) through (HC3), resulting in an increase in the areal fractions of surviving thru-holes in (HC4) and (HC5), even though the areal fraction of holes generated in each layer was the same at $a=0.1$ for all hole configurations. This trend was even more pronounced for (HC6) and (HC7), where holes were generated with open hexagonal cells selected with unequal probabilities of $P_i\;(i=1,2,3)$ to impart a certain directionality, as described in Sect.~\ref{method} and Fig.~\ref{size-of-hole}(c), which increased the shape anisotropy or aspect ratio. From Fig.~\ref{simulationI}(c), it can be seen that significantly more thru-holes survive in (HC6) and (HC7) compared to in (HC1)$\sim$(HC5). In fact, the surviving thru-holes in (HC6) and (HC7) were estimated to have $A_\mathrm{th}^\mathrm{(HC6)}=0.0583$ and $A_\mathrm{th}^\mathrm{(HC7)}=0.0587$, respectively. It was verified that the shape anisotropy enhanced by directionality allows more thru-holes to survive. Therefore, we confirm that naturally occurring linear-like cracks play an important role in the formation of thru-holes.

\begin{figure}
\centering
\includegraphics[width=1.0\columnwidth]{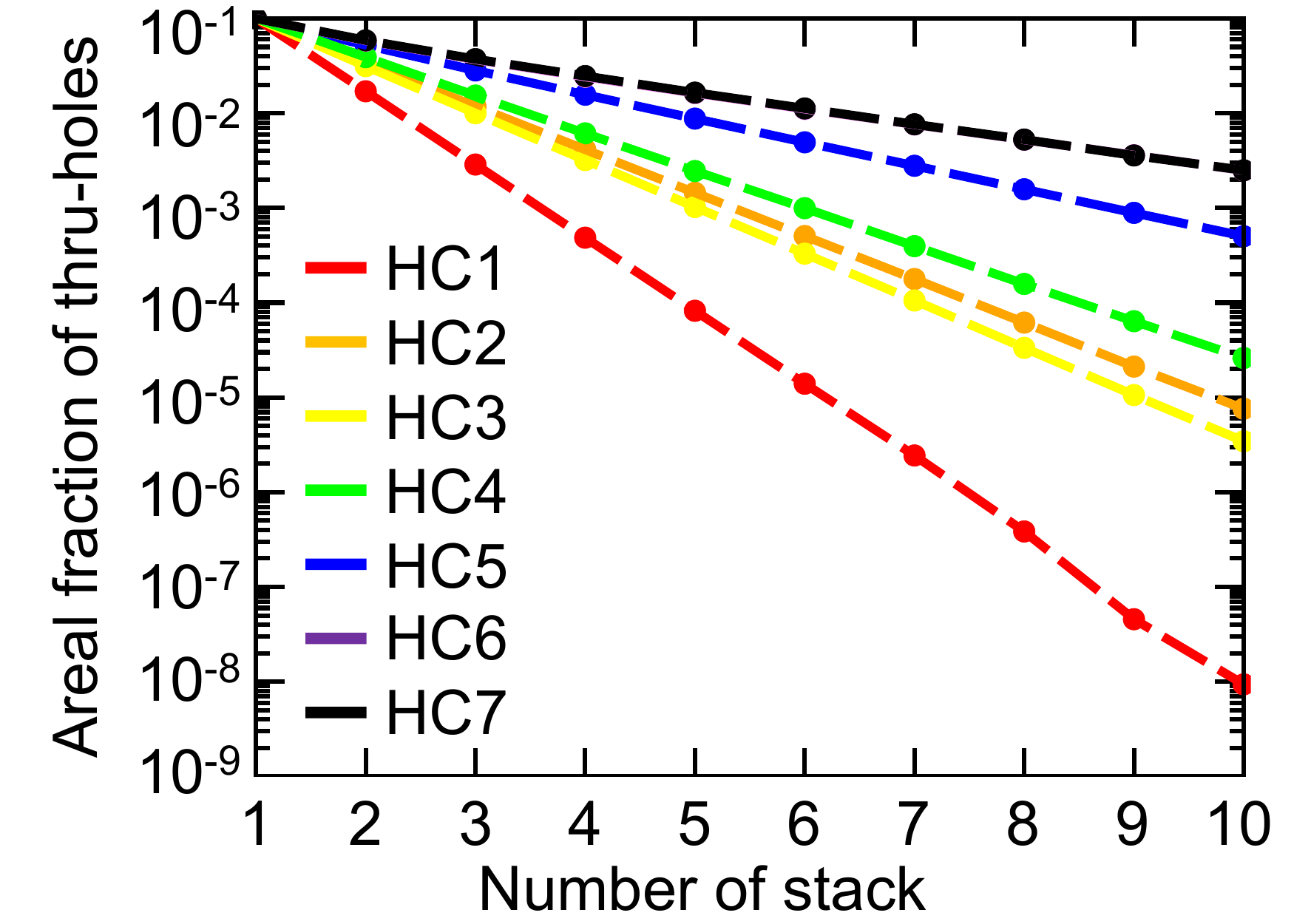}
\caption{\label{areal_ratio} Areal fraction of surviving thru-holes as a function of the number of stacks or layers for seven different hole configurations (HC1)$\sim$(HC7) described in Fig.~\ref{simulationI} with the same areal fraction of holes of $a=10$\% in each stack. Red, orange, yellow, green, blue, purple, and black lines represent the seven different hole configurations as indicated in the legend. The circular dots are the simulation data. Note that the purple curve for (HC6) is barely visible because it is drawn just below the black curve for (HC7).}
\end{figure}

We then ran our Monte Carlo simulation to explore how holes can survive as thru-holes during multiple stacking for the seven different hole configurations. The areal fraction of the surviving thru-holes $A_\mathrm{th}$ was evaluated as a function of the number of stacks or layers $n$. Figure~\ref{areal_ratio} shows the $A_\mathrm{th}$ evaluated for different hole configurations, all of which have the same areal fraction of holes of $a=10$\% in each stack. In each case, it decreases exponentially with the number of stacks, but with a different slope. In the case of (HC1), which is the hole configuration consisting essentially of isolated minimum size holes or hexagonal unit cells, it was expected that $A_\mathrm{th}$ would decrease as $r^n$ with a survival rate of $r=0.1$, which is the same as $a$, every stack. In our real simulation, however, the survival rate was estimated to be $r=0.17$ from Fig.~\ref{areal_ratio}, because there were holes larger than a single hexagonal cell even for (HC1). Nevertheless, this survival rate is still small enough that $A_\mathrm{th}$ would quickly decrease during multiple stacking. For the other hole configurations (HC2)$\sim$(HC7), $A_\mathrm{th}$ decreases much more slowly than for (HC1), similar to the case of two stacks shown in Fig.~\ref{simulationI}. Surprisingly, for the case of ten stacks, $A_\mathrm{th}$ for (HC7) is five orders of magnitude higher than that for (HC1). This result strongly supports that the survival probability of holes as thru-holes is much greater when the shape of the holes is much more anisotropic than when it is isotropic.

\begin{figure}
\centering
\includegraphics[width=1.0\columnwidth]{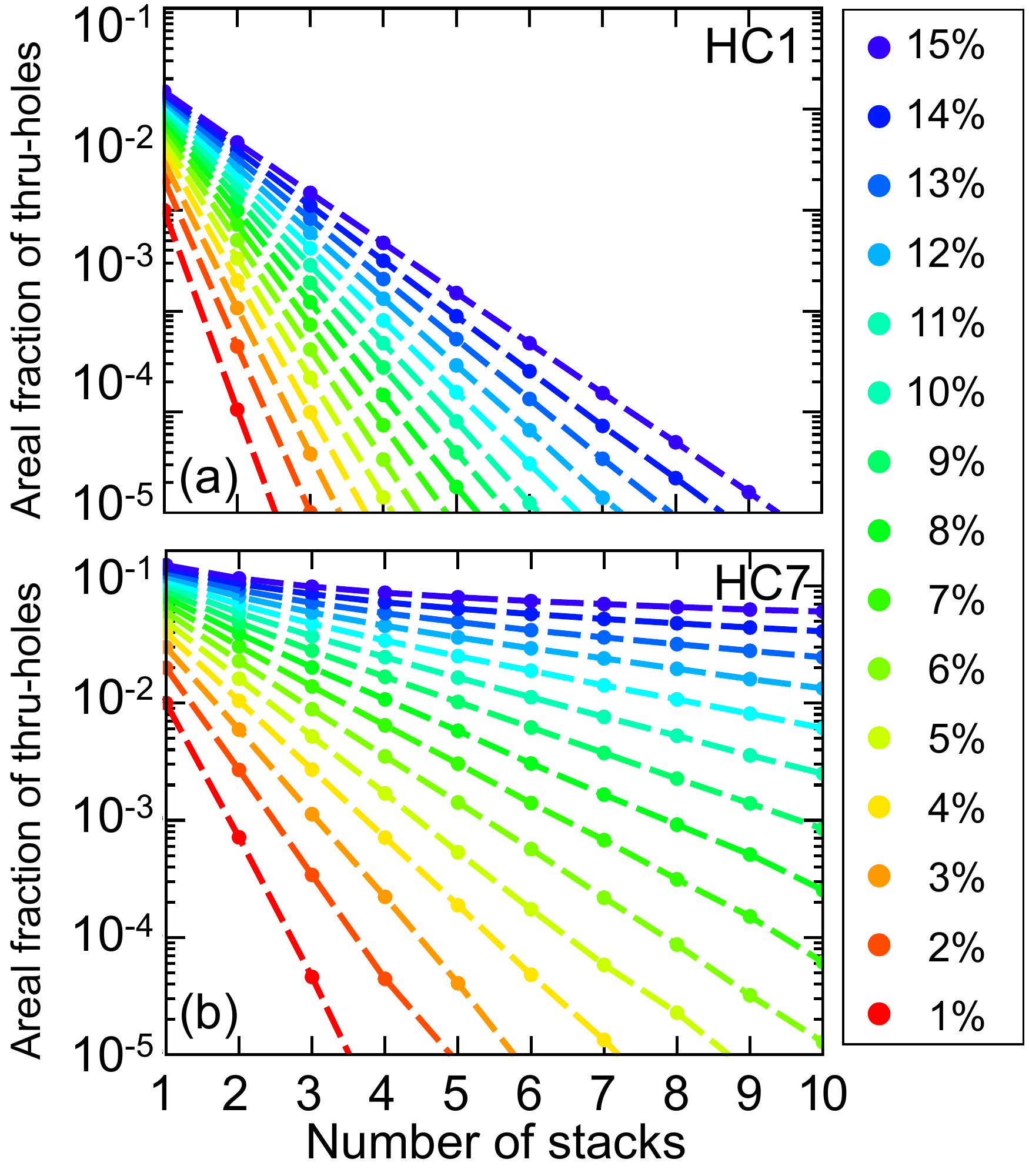}
\caption{\label{effect_of_areal_fraction_of_holes}Areal fraction of surviving thru-holes $A_\mathrm{th}$ as a function of the number of stacks $n$ for (a) (HC1) and (b) (HC7) while increasing the areal fraction of holes $a$ created in each stack by 1\% from 1\% to 15\%. A rainbow-colored scale was used to represent the $a$ values in each stack, with 1\% being red and 15\% being purple, as indicated in the legend on the right.}
\end{figure}


So far, we have kept $a=10$\% constant. To further investigate the effect of $a$ on thru-hole survival during multiple stacks, we also performed our Monte Carlo simulations while changing $a$. Overall, our simulation revealed that the higher the $a$, the higher the $A_\mathrm{th}$, as expected. Figure~\ref{effect_of_areal_fraction_of_holes} shows $A_\mathrm{th}$ as a function of $n$ while increasing $a$ from 1\% to 15\% for only two cases of (HC1) and (HC7) for comparison, since these two hole configurations yielded the smallest and the largest thru-hole survival probabilities at $a=10$\%, respectively. For both (HC1) and (HC7), it is shown that $A_\mathrm{th}$ on logarithmic scale decreases linearly with the number of stacks $n$ for every $a$ value. So these data were fitted with $A_\mathrm{th}=A_0r^n$, where $A_0$ is a proportional constant and $r$ is the survival rate. For example, with $a=1$\% and 15\%, we obtained $r=0.011$ and $0.313$ for (HC1) and $r=0.072$ and $0.897$ for (HC7), respectively. In other words, the survival rate of thru-holes strongly depends not only on the hole configuration, but also on the areal fraction of holes $a$ created in each stack for a given hole configuration, as shown in Fig.~\ref{effect_of_areal_fraction_of_holes}. Thus, this dependence can be used to estimate the areal fraction of holes in each stack as explained in the following paragraph.

\begin{figure}
\centering
\includegraphics[width=1.0\columnwidth]{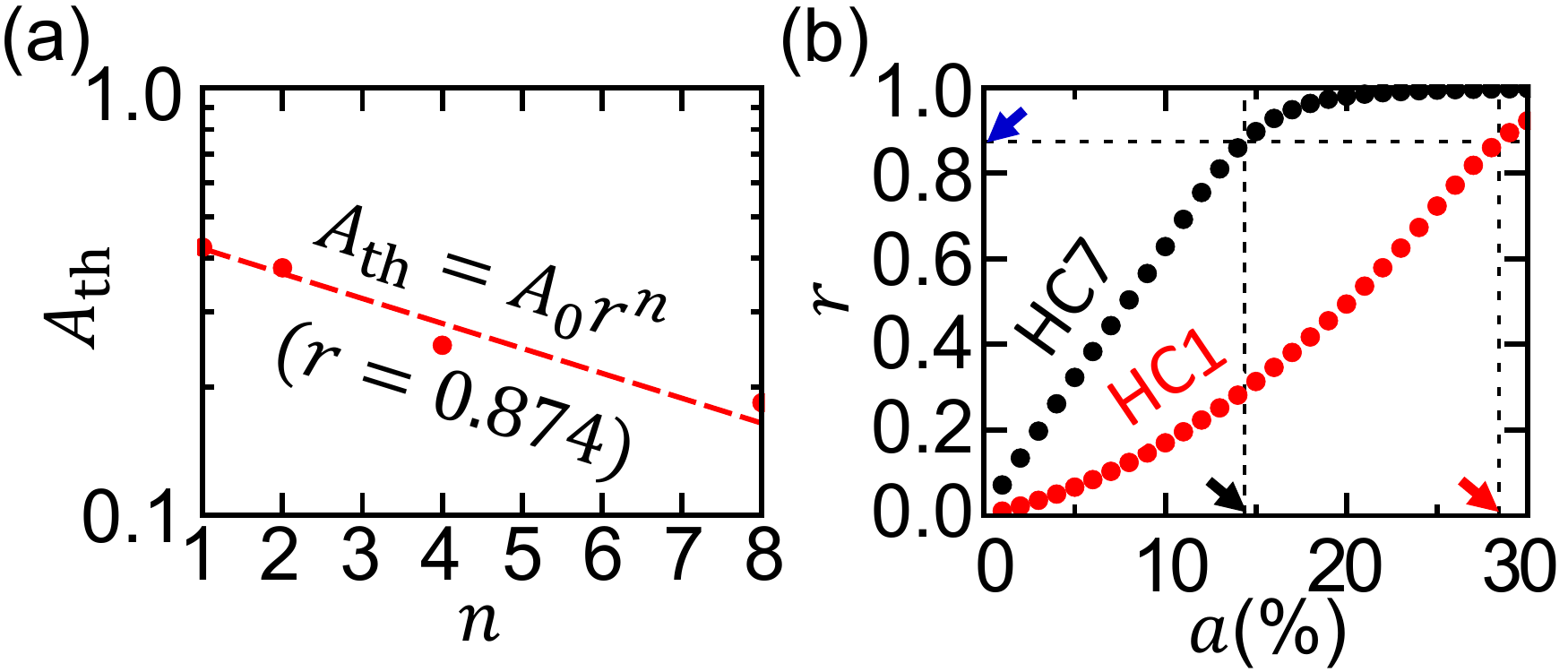}
\caption{\label{areal_ratio_of_GaN_domains}(a) Areal fraction of thru-holes $A_\mathrm{th}$ as a function of stacks $n$, estimated from our previous experimental data of grown GaN domains\cite{thruhole} and fitted with a fitting function of $A_\mathrm{th}=A_0r^n$, yielding $r$=0.874. (b) Thru-hole survival rate $r$ plotted as a function of the areal fraction of holes $a$ for configurations (HC1) and (HC7). The blue arrow with a horizontal dashed line indicates the survival rate of $r=0.874$ estimated in (a), and the red and black arrows with vertical dashed lines indicate the areal fraction of holes $a=28.4$\% and 14.4\% for (HC1) and (HC7), respectively, to realize the experimental result.}
\end{figure}


Based on the results of our computational simulations described above, we tried to estimate what the areal fraction of holes in each stack of $h$-BN would be in a real experimental situation. In our previous work on thru-hole epitaxy of GaN over multi-stacked $h$-BN on sapphire, the areal fraction of GaN domains grown over $h$-BN stacks decreased as the number of $h$-BN stacks increased.\cite{thruhole} Note that the areal fraction of GaN domains is closely related to the areal fraction of surviving thru-holes, but not exactly the same, because GaN domains were grown laterally over the thru-holes. However, the survival rates of the GaN domains and thru-holes can be considered essentially the same. From our previous experimental data,\cite{thruhole} we evaluated the areal fraction of GaN domains grown over $h$-BN stacked 1,2,4, and 8 times and obtained $r=0.874$ by fitting the evaluated data to $A_\mathrm{th}=A_0r^n$, as shown in Fig.~\ref{areal_ratio_of_GaN_domains}(a). We then estimated $a$, the areal fraction of holes created in each stack for (HC1) and (HC7), yielding this $r$ value, which corresponds to the thru-hole survival rate per stack in our experiment. Monte Carlo simulations were performed to generate $A_\mathrm{th}$ as a function of $n$ in the extended range of $a$ up to 30\% in addition to the data sets shown in Fig.~\ref{effect_of_areal_fraction_of_holes}. For each $a$ value, we obtained the survival rate $r$ by fitting the data set of $(n, A_\mathrm{th})$ to $A_\mathrm{th}=A_0r^n$, as shown in Fig.~\ref{areal_ratio_of_GaN_domains}(b). For the given value of $r=0.874$ marked by the blue arrow with a horizontal dashed line, the areal fraction of holes $a$ in each stack was estimated to be $r=28.4$\% and 14.4\% for (HC1) and (HC7) marked by the red and black arrows with vertical dashed lines, respectively. This indicates that the thru-hole survival rate $r$ is not equal to the areal fraction of holes $a$, unlike the ideal case where $r=a$. More surprisingly, $r$ can even quickly approach 1 as $a$ increases, especially in certain configurations where the holes are highly anisotropic in shape, such as (HC7), as shown in Fig.~\ref{areal_ratio_of_GaN_domains}(b).

%


\section{Summary and Conclusions}
We investigated the effect of hole configuration on the survival of holes as thru-holes by carrying out Monte Carlo simulations in various hole configurations. The survival rate of holes as thru-holes was found to be critically dependent on both the shape of the holes and their areal fraction in each stack. In particular, we showed that much less areal fraction of holes is required to produce a given level of the areal fraction of thru-holes than intuitively estimated when the holes are highly anisotropic in shape with a moderate areal fraction of holes in each stack. Our simulations validate thru-hole epitaxy by showing that it does not require an unreasonably large areal fraction of holes in each stack to maintain the areal fraction of thru-holes in multiple stacks as long as the holes in each stack are highly anisotropic in shape.


\section*{Acknowledgements}
We acknowledge financial support from the Korean government through the National Research Foundation of Korea (NRF-2022R1A2C1005505, NRF-2022M3F3A2A01073562, NRF-2019R1F1A1063643, NRF-2021R1A5A1032996, RS-2023-00240724) and the Brain Korea 21 Four Program in 2021. The computational work was done using the resources of the KISTI Supercomputing Center (KSC-2022-CRE-0379, KSC-2023-CRE-0053).

\section*{Conflict of Interest} \par
The authors declare no conflict of interest.

\section*{Data Availability Statement} \par
The data that support the ﬁndings of this study are available from the corresponding author upon reasonable request.


\end{document}